# Alloy Engineering of Polar $(Si,Ge)_2N_2O$ System for Controllable Second Harmonic Performance

Lei Kang,[1,2] Gang He,[2] Xinyuan Zhang,[2] Jiangtao Li,[2] Zheshuai Lin,[2]* Bing Huang[1]†

[1] *Beijing Computational Science Research Center, Beijing 100193, China*
[2] *Technical Institute of Physics and Chemistry, Chinese Academy of Sciences, Beijing 100190, China*
[3] *Institute of Functional Crystals, Tianjin University of Technology, Tianjin 300384, China*
E-mails: zslin@mail.ipc.ac.cn (Z.L.); bing.huang@csrc.ac.cn (B.H.)

**Although silicon oxynitrides are important semiconductors for many practical applications, their potential second-order nonlinear optical (NLO) applications, regardless of balanced or controllable performance, have never been systemically explored. Using the first-principles calculations, in this article, we discover that the sinoite (i.e., typical silicon oxynitride $Si_2N_2O$) can simultaneously exhibit wide optical bandgap, strong second-harmonic generation (SHG) effect, and large birefringence, which are further confirmed by our preliminary experimental data. Importantly, we propose that alloying engineering can be further applied to control the balanced NLO properties in the $Si_2N_2O$ system. Combining first-principles calculations and cluster expansion theory, we demonstrate that alloying Ge into $Si_2N_2O$ can easily form low formation energy $Si_{2(1-x)}Ge_{2x}N_2O$ alloys, which can in turn achieve controllable phase-matching harmonic output with high SHG efficiency at different energy ranges. Therefore, alloy engineering could provide a unique approach to effectively control the balanced NLO performance of $Si_{2(1-x)}Ge_{2x}N_2O$, making this polar alloy system holding potential applications in tunable laser frequency conversion and controllable all-optical devices.**

## 1. Introduction

Silicon nitrides and oxynitrides are well-known multi-functional material systems that play an important role in a wide range of applications, *e.g.*, ceramic bearings, cutting tools, electronics, high-temperature materials, and ultraviolet (UV) light-emitting diodes, due to their good performance in hardness, thermal stability, wear resistance and optical bandgap [1-9]. To further extend their applications to laser technology, optical precision measurement, quantum information and computing [10-13], it is highly desirable to realize nonlinear optical (NLO) frequency conversion in these systems. Unfortunately, to best of our knowledge, the feasibility of NLO properties in these systems, *e.g.*, second harmonic performance, regardless of balanced or controllable, remains unclear due to the lack of a systematic understanding.

In general, for a NLO system (here mainly refers to the second-order nonlinearity), the ideal situation is to simultaneously realize wide energy bandgap ($E_g$), strong second harmonic generation (SHG) effect ($d_{ij}$), and large optical birefringence ($\Delta n$), which are three critical but competing factors in a NLO system [14]. It is known that these three factors difficult to be achieved in one single material system. For examples, a larger $E_g$ is usually accompanied by a smaller first-order (*e.g.*, refractive indices $n_i$) and second-order (*e.g.*, $d_{ij}$) optical susceptibility; a smaller $n_i$ and lower optical anisotropy can result in a smaller $\Delta n$. In practice, the UV NLO materials, *e.g.*, β-$BaB_2O_4$ (BBO) and $LiB_3O_5$ (LBO) [15,16], usually have relatively large $E_g$ but small $d_{ij}$. Therefore, although they can meet the basic requirements of NLO application in the UV region (*e.g.*, 355 nm for third harmonic generation of the pratical Nd: YAG 1064 nm lasers), they cannot realize efficient NLO conversion in the infrared (IR) region (*e.g.*, 4 μm for the pratical atmospheric transparent window) due to the small $d_{ij}$. In comparison, the IR NLO materials, *e.g.*, $AgGaS_2$ and $ZnGeP_2$ [17,18], usually have relatively large $d_{ij}$ but small $E_g$. Although they can meet the IR NLO conversion with certain efficiency, they are opaque to the UV light due to the small $E_g$. Therefore, a good balance among the $E_g$, $d_{ij}$, and $\Delta n$ is critical for achieving the balanced NLO performance in different energy regions [10], which, however, is challenging to be realized due to the lack of a general design principle.

Based on our physical intuition, we expect that the silicon nitride/oxynitride may be ideal material platforms that can achieve balanced NLO performance in wide spectra from UV to IR, in terms of their following advantages: (**i**) wide $E_g$ for UV optical transparency, (**ii**) high mechanical-thermal stability and laser-induced damage threshold due to the existence of Si-N or/and Si-O tetrahedra, (**iii**) non-toxic, and (**iv**) stable electronic property, high optical efficiency, insensitive to temperature and drive current [1-9]. More importantly, the silicon nitride/oxynitride may easily form solid solutions with their germanium (Ge) analogue, due to their similar structural and chemical properties. Therefore, alloy engineering may provide a unique approach to tune their overall linear and NLO properties, *e.g.*, $E_g$, $\Delta n$ and $d_{ij}$, which can have various important applications especially in terms of performance regulation [19-25]. In view of these superior properties, it is highly desirable to evaluate the potential linear and NLO performance in the silicon nitrides/oxynitrides and their potential alloys, *e.g.*, for the SHG process applied to pratical frequency conversion.

In this article, we propose a general design principle for achieving the good balance of NLO performance by alloy engineering, which is successfully demonstrated in the silicon oxynitride system. First, using the first-principles calculations, we discover that the $Si_2N_2O$ can simultaneously exhibit wide optical $E_g$ (~5 eV), strong SHG $d_{ij}$ effect (~2.6×KDP), and large $\Delta n$ (~0.08), which are further confirmed by our preliminary experimental measurements. Second, combining first-principles calculations and cluster expansion theory, we demonstrate that alloying Ge into $Si_2N_2O$ can easily form low formation energy $Si_{2(1-x)}Ge_{2x}N_2O$, which can in turn achieve controllable phase-matching harmonic output with high SHG efficiency at different energy ranges. Therefore, alloy engineering indeed can provide a unique approach to effectively control the overall NLO performance of $Si_{2(1-x)}Ge_{2x}N_2O$, making it useful for different energy-range applications with well-balanced NLO performance.

## 2. Design Principle of Alloy Engineering for Balanced NLO Performance

As an important criterion for practical NLO materials, the NLO conversion efficiency in a phase matchable SHG process can be evaluated according to the following formula under the non-depletion approximation:

$$\eta = \frac{8\pi^2 d_{eff}^2 \cdot L^2 \cdot I_\omega}{\varepsilon_0 c \cdot n_\omega^2 n_{2\omega} \lambda_\omega^2} \quad (1)$$

where $\eta$ is the conversion efficiency, $d_{eff}$ is the effective SHG coefficient (generally proportional to $d_{ij}$ along the PM direction), $L$ is the length of a crystal, $I_\omega$ is the peak power density of the input beam, $\varepsilon_0$ is the vacuum permittivity, $n_\omega$ and $n_{2\omega}$ are the refractive indices at the fundamental $\lambda_\omega$ and the second harmonic $\lambda_{2\omega}$ along the PM direction, and $c$ the light speed in vacuum.

According to Eq. (1), in order to achieve a similar efficiency $\eta$, the NLO materials at different energy region should exhibit $\lambda_\omega$-dependent $d_{eff}$. For example, for the 4μm IR light, in order to achieve the same efficiency $\eta$ as the 400nm UV light, under the same conditions, the $d_{eff}$ of an IR NLO material, *e.g.*, the widely used all-round NLO crystal $KTiOPO_4$ (KTP) with $d_{eff}$ = 3.2 pm/V [26], needs to be ~10 times larger than that of an UV NLO material, *e.g.*, the important benchmark NLO material $KH_2PO_4$ (KDP) with $d_{36}$ = 0.39 pm/V. Therefore, we can understand that the BBO and LBO cannot achieve the efficient frequency conversion in the IR region, because their SHG effects are only around 3-5×KDP [10].

Accordingly, a good UV NLO material should satisfy the following balanced linear and NLO performance, including an energy bandgap $E_g \geq 4.7$ eV for short UV absorption edge $\lambda_{UV} \geq 266$ nm, an SHG coefficient $d_{ij} \geq 1\times$KDP, and a birefringence $\Delta n \geq 0.08$ (approximately) to realize moderate PM output of 355 nm and 266 nm, respectively, corresponding to the third and fourth harmonic generation of the pratical Nd: YAG 1064 nm lasers. Correspondingly, a good IR NLO material needs a $E_g \geq 3$ eV (mainly to avoid multiphoton absorption damage for the μm-lasers), a $d_{ij} \geq 10\times$KDP (or 1×KTP), and a $\Delta n \geq 0.05$ (approximately).

To achieve the balanced requirements for potential NLO applications in both UV and IR regions, we take the silicon nitride/oxynitride system as an example to illustrate our design principles by alloy engineering, which is generally applicable to other similar systems. Note that, for most of silicon nitride compounds, based on their basic $(SiN_4)^{8-}$ tetrahedral motifs as schematically shown in **Figure 1a**, they usually have large $E_g$ due to strong $sp^3$-hybrid Si-N bond, but small structural and optical anisotropy (*i.e.*, small $\Delta n$ and $d_{ij}$) due to regular tetrahedral coordination. As a result, the balance between these three key NLO parameters (*i.e.*, $E_g$, $\Delta n$ and $d_{ij}$) cannot be achieved, preventing them to meet the PM requirements for the SHG conversion especially in the UV region [27].

Here, we propose two feasible steps to enhance the $\Delta n$ and $d_{ij}$. First, to enhance the $\Delta n$, based on the anionic group theory [10], one can introduce some anionic components into the $(SiN_4)^{8-}$ tetrahedra to enlarge the structural anisotropy. As shown in **Figure 1b**, an effective way is to replace the bridge-site nitrogen with oxygen forming the oxynitride structure. In this case, the $SiN_3$-O-$SiN_3$ structure is built by distorted $(SiN_3O)^{7-}$ tetrahedra connected through O atoms along the polar axis, which may exhibit similar $E_g$ and $d_{ij}$ but larger $\Delta n$ as compared with $(SiN_4)^{8-}$ in **Figure 1a**. Second, in order to further enhance the $d_{ij}$, alloying Ge into $(SiN_3O)^{7-}$ forming ordered $[(Si,Ge)N_3O]^{7-}$ solid solution, as illustrated in **Figure 1c**, could lower the structural symmetry and enhance the NLO response. Meanwhile, alloy engineering may also be used to tune the $E_g$ for various NLO applications in different energy ranges, due to the different $E_g$ of $(SiN_3O)^{7-}$ and $(GeN_3O)^{7-}$ hosts, providing an effective way to control the overall linear and NLO performance.

## 3. Computational and Experimental Methods

To verify the design principle shown in **Figure 1**, based on the first-principles calculations, we have evaluated the linear and NLO properties of the corresponding material systems, including typical silicon nitrides, silicon oxynitrides, and germanium oxynitrides as listed in **Table 1**. The first-principles calculations are performed by the plane-wave pseudopotential method using CASTEP based on the density functional theory (DFT) [28]. The norm-conserving pseudopotentials and generalized gradient approximation (*e.g.*, Perdew-Burke-Ernzerhof functional) are employed [29,30]. The lattice constants and atomic positions are optimized using the quasi-Newton method [31]. The electronic structures, linear and NLO properties are obtained according to our proposed methods [32,33].

In order to effectively study the $(Si,Ge)_2N_2O$ binary alloy systems, the first-principles based cluster expansion (CE) approach, as implemented in ATAT code [34], is employed to generate the random $Si_{2(1-x)}Ge_{2x}N_2O$ alloy structures at different $x$ and simulate the formation energies of $Si_{2(1-x)}Ge_{2x}N_2O$ systems. See more details of the CE methods in the following discussions.

In addition, the powder X-ray diffraction (XRD), UV-visible-IR absorption spectrum and Kurtz-Perry powder SHG measurements are used to characterize the fundamental properties of synthetic $Si_2N_2O$ samples in the experiments, which is mainly used to confirm the accuracy of our computational methods. For more computational and experimental details, see **Methods** in the Supplemental Material [35].

## 4. Results and Discussion

***4.1. NLO properties of silicon nitrides and oxynitrides.*** Silicon nitride $\alpha$-$Si_3N_4$ [36], an important semiconductor system, is constructed by the trigonal lattice ($P3_1c$ symmetry) with corner-shared $(SiN_3)_3$-N tetrahedral motifs as shown in **Figure 2a**, same as the $SiN_3$-N-$SiN_3$ units shown in **Figure 1a**. The first-principles results show that $\alpha$-$Si_3N_4$ has a suitable (but not strong) SHG effect (~ 0.5 pm/V**,** see **Table 1**), in good agreement with the experimental data (~1×KDP for thin film) [37]. However, its $\Delta n$ is too small (< 0.02 at 1 μm) to achieve the important UV coherent output, *e.g.*, 355 nm, though it has a large $E_g$ ≈ 4.6 eV ($\lambda_{UV}$ ≈ 270 nm). It clearly indicates that $\alpha$-$Si_3N_4$ cannot achieve a good NLO balance among $E_g$, $d_{ij}$, and $\Delta n$ for UV SHG. Similarly, the alkali-metal silicon nitride $LiSi_2N_3$ also has a small $\Delta n$ (<0.03) [38], even its $E_g$ (> 5 eV) and $d_{ij}$ (≈ 5×KDP) are relatively large (**Table 1**).

As proposed in **Figure 1b**, the oxygen doping of silicon nitride forming distorted silicon oxynitride tetrahedra may enlarge the optical anisotropy of $Si_3N_4$. To confirm this, we select the $Si_2N_2O$ structure as the typical example to investigate its linear and NLO properties. As shown in **Figure 2b**, the $Si_2N_2O$ is composed by the $(SiN_3O)^{7-}$ tetrahedra connected through oxygen atoms along the *c*-axis, same as the case in **Figure 1b** [39-41]. Considering the larger electronegativity of O than N and the shorter bond length of Si-O (~1.6 Å) than Si-N (~1.8 Å), $Si_2N_2O$ can exhibit a wider $E_g$ (due to stronger bonding-antibonding separation) and a larger structural anisotropy (due to the tetrahedral distortion) than α-$Si_3N_4$. Indeed, the calculated $E_g$, $d_{ij}$, and $\Delta n$ shown in **Table 1** confirm our physical intuition. Comparing **Figures 2a** and **2b**, it is seen that the $Si_2N_2O$ exhibits a stronger $d_{ij}$ (~2.6×KDP) and a larger $\Delta n$ (~0.08) than those of $Si_3N_4$. Similar trend of enhanced $d_{ij}$ and $\Delta n$ are also observed in our study for the alkali-metal silicon oxynitride LiSiNO ($d_{ij} \approx 1.4$×KDP, $\Delta n \approx 0.04$, see **Table 1**) compared to those of $LiSi_2N_3$ [42].

***4.2. Experimental measurements on silicon oxynitride of $Si_2N_2O$.*** To further confirm our design strategy and theoretical prediction, we have performed the experimental measurements for the $Si_2N_2O$ system. The powder XRD results of synthetic $Si_2N_2O$ samples are shown in **Figure 3a**, which is consistent with the existing XRD data of $Si_2N_2O$. The UV-visible-IR transmission and absorption spectra are measured as shown in **Figure 3b**, indicating that the UV absorption edge $\lambda_{UV}$ is at ~230 nm, corresponding to the $E_g \approx 5.4$ eV, which is slightly larger than α-$Si_3N_4$ ($E_g \approx 5$ eV). The Kurtz-Perry powder SHG measurement of $Si_2N_2O$ samples is shown in **Figure 3c** [43]. Due to the extremely small powder-size (<10 μm, see **Figure S1** [35]) of our synthetic samples, the powder SHG signal with respect to the particle size cannot be characterized experimentally. However, compared with the SHG signal and trend of KDP benchmark, it is reasonable to believe that the SHG intensity of $Si_2N_2O$ should be larger than that of KDP ($d_{36}$ = 0.39 pm/V), which is qualitatively consistent with our theoretical prediction (~1 pm/V in **Table 1**). Meanwhile, according to the Kurtz-Perry derivation, the powder SHG effect can be obtained from the SHG coefficients [43]; the simulated $d_{powder}$ of $Si_2N_2O$ is 0.95 pm/V (2.4×KDP), comparable as the maximum SHG coefficient $d_{24}$ ~ 1 pm/V.

Although it is difficult to measure the refractive indices directly due to the limited size of $Si_2N_2O$ samples, its structural properties have been measured in other experiments [39,40]. The experimental mineral data shows that its biaxial refractive indices $n_x$ = 1.740, $n_y$ = 1.855 and $n_z$ = 1.855, which is in good agreement with our calculated results ($n_x \approx 1.777$, $n_y \approx 1.835$ and $n_z \approx 1.856$, relative error < ±2%). Based on the calculated dispersion curves of refractive indices, as plotted in **Figure 3d**, the $Si_2N_2O$ can achieve the shortest type-I PM output wavelength $\lambda_{PM} \approx 285$ nm, shorter than the important 355 nm in the UV region.

***4.3. Tunable NLO properties in alloying (Si,Ge)$_2$N$_2$O.*** As proposed in **Figure 1c**, it is highly possible to further enhance the SHG effects in the silicon oxynitride system by alloying to form (Si,Ge)$_2$N$_2$O solid solution [25,44]. Since $(GeN_3O)^{7-}$ has longer Ge-N/O bond length and larger tetrahedral distortion than those of Si-N/O in $(SiN_3O)^{7-}$, the structural anisotropy would be enlarged so that a sufficiently large birefringence can be maintained (see **Figure 2c**). In addition, it is expected that the $E_g$ of (Si,Ge)$_2$N$_2$O can be tuned in a wide range, depending on the Ge concentration, due to large $E_g$ difference between $Si_2N_2O$ (~5 eV) and $Ge_2N_2O$ (~2.8 eV). Therefore, it is expected that the $Si_{2(1-x)}Ge_{2x}N_2O$ can be applied to realize the NLO applications in different energy ranges.

In order to examine the feasibility of forming (Si,Ge)$_2$N$_2$O alloys, based on the first-principles based CE approach [34], we have systemically investigated the formation energies ($E_f$) of $Si_{2(1-x)}Ge_{2x}N_2O$ ($0 \le x \le 1$) binary alloys. Here, the $E_f$ is defined as

$$E_f = E\left[Si_{2(1-x)}Ge_{2x}N_2O\right] - (1-x)\mu_{Si_2N_2O} - x\mu_{Ge_2N_2O} \quad (2)$$

where $\mu_{Si2N2O}$ ($\mu_{Ge2N2O}$) is the total energy of $Si_2N_2O$ ($Ge_2N_2O$) per unit cell (see **Methods** [35] for more details). The basic idea of CE is to expand the energies of a $Si_{2(1-x)}Ge_{2x}N_2O$ configuration into energy contributions of cluster figures (single atoms, pairs, triples, *etc.*) based on a generalized Ising Hamiltonian:

$$E(\sigma) = J_0 + \sum_i J_i \hat{S}_i(\sigma) + \sum_{i<j} J_{ij} \hat{S}_i(\sigma)\hat{S}_j(\sigma) + \sum_{i<j<k} J_{ijk} \hat{S}_i(\sigma)\hat{S}_j(\sigma)\hat{S}_k(\sigma) + \cdots \quad (3)$$

In Eq. (3), the index *i*, *j*, and *k* run over all the alloy sites, and $S_m(\sigma)$ is set to +1 (-1) when it is occupied by $Si_2N_2O$ ($Ge_2N_2O$) dimer. Ideally, the CE can represent any $Si_{2(1-x)}Ge_{2x}N_2O$ alloy energy $E(\sigma)$ by the appropriate selection of *J*, which can be fitted from the first-principles total energy calculations based on a sufficient number of random alloy configurations. To calculate the binary phase diagram of $Si_{2(1-x)}Ge_{2x}N_2O$, the Monte Carlo simulations, which sample a semi-grand-canonical ensemble, are carried out in which the energetics of $Si_{2(1-x)}Ge_{2x}N_2O$ are specified by the CE Hamiltonian.

First, we have calculated the $E_f$ of 119 $Si_{2(1-x)}Ge_{2x}N_2O$ alloys (<20 atoms/supercell) with different structural symmetries at different Ge concentration ($x$) based on the first-principle methods. Second, the self-consistent CE fitting is employed to fit the DFT-calculated $E_f$, to obtain the most important effective cluster interactions *J*. Our calculations confirm that these 119 alloy structures are sufficient to obtain the converged $J$ for the CE Hamiltonian. Therefore, the CE Hamiltonian of the binary alloy $Si_{2(1-x)}Ge_{2x}N_2O$ can be well obtained. Third, we can use this Hamiltonian to quickly predict the $E_f$ of all the nonequivalent alloy configurations. The DFT-calculated, CE-fit and CE-predicted $E_f$ are shown in **Figure 4a**. Interestingly, it is found that the $E_f$ of all the $Si_{2(1-x)}Ge_{2x}N_2O$ alloys are rather low, mostly in the range between 0 and 40 meV, indicating that they are easily to be formed at finite temperature. Indeed, by employing the CE-based MC simulations, we further obtained the $T$-$x$ phase diagram of $Si_{2(1-x)}Ge_{2x}N_2O$ as a function of $x$. As shown in the inset of **Figure 4a**, the miscibility gap of $Si_{2(1-x)}Ge_{2x}N_2O$ is ~155 K, which is even below the room temperature. Note that this miscibility gap is obtained based on the thermal equilibrium conditions, assuming that the alloyed atoms are free to move to reach their equilibrium positions. In practice, the sizable diffusion barriers need to be overcome for the atoms to move in $Si_{2(1-x)}Ge_{2x}N_2O$.

Interestingly, we notice that there is an existing ordered $SiGeN_2O$ phase with the *Cm* symmetry in the crystal structure database (see **Figure 2d**) [45,46]. For comparison, its calculated $E_f$ (marked as red star) is also shown in **Figure 4a**. Obviously, there are a couple of $SiGeN_2O$ structures predicted by our CE calculations can have even lower $E_f$ than that of the *Cm*-phase of $SiGeN_2O$, *e.g.*, the *Cc*- and $P2_1$-phase $SiGeN_2O$ (for details see **Figure S2** [35]), indicating that many $SiGeN_2O$ structures may also be able to be synthesized in the experiments.

Taking the $Si_{2(1-x)}Ge_{2x}N_2O$ ($0 \leq x \leq 1$) structures with lowest $E_f$ in **Figure 4a** at different $x$ as examples, their linear and NLO properties can be calculated, as shown in **Figure 4b** (for details see **Table S1** [35]). Again, the calculated linear and NLO properties of the *Cm*-phase of $SiGeN_2O$ are also listed in **Table 1** for comparison. It is seen that when $x$ increases from 0 to 1, the SHG intensity $d_{ij}$ of $Si_{2(1-x)}Ge_{2x}N_2O$ can gradually increases from ~2.6×KDP to ~19×KDP, validate our alloy design principle shown in **Figure 1c**. Meanwhile, the $\Delta n$ can maintain sufficiently large values between 0.07 and 0.10, which are good enough for most PM processes from UV to IR. However, the $E_g$ smoothly decreases as $x$ increases, as expected (**Table 1**). Therefore, the PM wavelength $\lambda_{PM}$ is red shifted from 285 to 444 nm.

In principle, the $Ge_2N_2O$ cannot achieve the UV laser generation due to its $\lambda_{PM} \approx 444$ nm, despite it exhibits larger SHG effect than KTP [26]. In contrast, the $SiGeN_2O$ has a larger $E_g \approx 3.6$ eV that can achieve the 355 nm output ($\lambda_{PM} \approx 345$ nm), satisfying a good balance of UV NLO performance (**Figure 2d**). Importantly, in this region, $SiGeN_2O$ exhibits stronger SHG effect (≈ 12×KDP, **Table 1**) than many practical UV NLO materials (*e.g.*, ~3-5×KDP for LBO and BBO). Furthermore, when $x$ is tuned in the range of 0.5~0.75, the $Si_{2(1-x)}Ge_{2x}N_2O$ can exhibit relatively large $E_g$ (>3 eV) and strong SHG $d_{ij}$ (>10×KDP), satisfying the balanced IR NLO requirement [47], as shown in **Figure 4c**. It indicates that they could be available for possible IR NLO conversion. All the results demonstrate that the polar (Si,Ge)$N_2O$ system can have strong NLO effects in wide energy regions, exhibiting either larger $d_{ij}$ (*e.g.*, than BBO at 355 nm) or lager $E_g$ (*e.g.*, than $AgGaS_2$ at 1 μm) respectively for UV or IR SHG conversion.

***4.4. SHG origin in $Si_2N_2O$ and $SiGeN_2O$***. To further understand the underlying NLO mechanism, the electronic band structures, partial density of states (PDOS), band resolved SHG analysis and SHG weighted density of $Si_2N_2O$ and $SiGeN_2O$ are calculated and plotted in **Figure 5** (see **Methods** [35] for more details). Because of the direct $E_g$ shown in **Figures 5a** and **5d**, $Si_2N_2O$ and $SiGeN_2O$ are superior to the semiconductors with indirect-$E_g$ in terms of the optical transition properties. These optical transitions are mainly originated between the *p* orbitals of N/O and *s* orbitals of Si/Ge around the band edges (**Figures 5b** and **5e**). Since the 2*p* orbitals of N/O contribute significantly on the valence band maximum (VBM), these N/O orbitals can contribute dominantly to the SHG effects, exhibiting strong SHG densities in $Si_2N_2O$ and $SiGeN_2O$, as shown in **Figure 5c** and **5f** [48].

Furthermore, comparing to $Si_2N_2O$, due to the inclusion of Ge, the conduction band minimum (CBM) of $SiGeN_2O$ is mainly contributed by the 4*s* orbitals of Ge, due to the lower orbital energy of Ge 4*s* than Si 3*s*. Meanwhile, the VBM maintains to be contributed by the orbitals of N/O (**Figure 5e**). Consequently, the electronic states on the Ge site exhibit more dominant SHG contribution and densities than that on the Si site (**Figure 5f**). Consequently, the $SiGeN_2O$ can exhibit a stronger SHG effect (~12×KDP) than $Si_2N_2O$ (~2.6×KDP). The results are overall consistent with our design principle explained in **Figure 1**.

## 5. Outlook and Conclusion

Before conclusion, we will give an outlook on the possible NLO applications of the polar $(Si,Ge)_2N_2O$ system. On the one hand, if the large-size crystals of $(Si,Ge)_2N_2O$ can be obtained, they can be directly used in traditional NLO devices to achieve tunable SHG PM output. In addition, they might possess favorable machining performance, due to good mechanical properties (*e.g.*, large bulk modulus ~ 155 GPa for $Si_2N_2O$). In practice, the strong covalent bonding of $Si_2N_2O$ results in high flexural strength and resistance to heating and oxidation up to temperatures of about 1350°C [49]. Moreover, $Si_2N_2O$ can be easily obtained from the combustion synthesis method of $3Si + 2N_2 + SiO_2 \rightarrow 2Si_2N_2O$. Although there are many studies on the functionalities of $Si_2N_2O$ [1-9], its NLO properties are presented for the first time in this study. Therefore, the further experimental synthesis of large-size crystal is of great importance to evaluate its NLO performance.

On the other hand, if the large-size $(Si,Ge)_2N_2O$ crystals are difficult to be obtained in the experiments, they can also be made into the pole glass, similar to the pole $SiO_2$ glass for possible electro-optic applications. For example, the thermal poling, which was electrothermal in nature, has been demonstrated in bulk $SiO_2$ samples by Myers et al. [50] The nonlinearity induced, which was evaluated from the SHG signals of 1064 nm laser, is of a similar magnitude to that in quartz. The related analysis indicates that in so-called glass ceramics, embedded crystals seem to be responsible for prominent SHG effect. Therefore, the further investigation of the NLO properties of the pole $(Si,Ge)_2N_2O$ glass states by both theories and experiments is highly desired.

In conclusion, we have proposed a general proposal to realize the balanced and tunable NLO properties of semiconductors by alloy engineering, which has been well demonstrated in the $(Si,Ge)N_2O$ system. The polar $(Si,Ge)N_2O$ system can achieve tunable energy bandgaps and PM SHG outputs from the UV to IR region. As a representative, $SiGeN_2O$ is predicted to exhibit sufficiently large SHG effect and birefringence for NLO crystals. We believe that this study may re-arouse the interests to theorists and experimentalists in the classical silicon oxynitride system for broad NLO applications.

**Acknowledgements:** The authors thank P. Gong for helpful discussion. This work was supported by the Science Challenge Project (Grant No. TZ2016003), NSFC (Grants Nos. 11634003, 51872297, 11704023, 51890864), and NSAF (Grants No. U1930402).

Table 1. Linear and NLO Properties of Typical Silicon Nitrides, Silicon Oxynitrides and Germanium Oxynitrides

| Symmetry | | | NLO-active unit | $\lambda_{UV}$ (nm) | $E_g$ (eV) | $d_{ij}$ (pm/V) | $\|d_{ij}\|_{max}$/ $d_{36}$(KDP) [c)] | $\Delta n$ at 1 µm |
|---|---|---|---|---|---|---|---|---|
| α-$Si_3N_4$ | *P3₁c* | exp. | $(SiN_3)_3$-N | 249 | 5.0 | ~ KDP | 1.0 | --- |
| | | cal. [a)] | | 270 | 4.6 | $d_{15}$=0.50; $d_{24}$=0.50; $d_{33}$=-0.44 | 1.3 | 0.016 |
| $LiSi_2N_3$ | $Cmc2_1$ | cal. [a)] | $(SiN_3)_2$-N | 249 | 5.0 | $d_{15}$=2.30; $d_{24}$=0.91; $d_{33}$=-0.21 | 5.9 | 0.029 |
| $Si_2N_2O$ | $Cmc2_1$ | exp. | $(SiN_3)_2$-O | 230 | 5.4 | $d_{powder}$ > KDP | | 0.115 |
| | | cal. [a)] | | 249 | 5.0 | $d_{15}$=0.46; $d_{24}$=1.00; $d_{33}$=-0.86 | 2.6 | 0.080 |
| LiSiNO | $Pca2_1$ | cal. [a)] | $(SiN_2O)_3$-N | 230 | 5.4 | $d_{15}$=0.37; $d_{24}$=0.56; $d_{33}$=0.10 | 1.4 | 0.040 |
| $Ge_2N_2O$ | $Cmc2_1$ | cal. [a) b)] | $(GeN_3)_2$-O | 444 | 2.8 | $d_{15}$=3.41; $d_{24}$=1.13; $d_{33}$=-7.46 | 19.1 | 0.091 |
| KGeNO | $Pca2_1$ | cal. [a)] | $(GeN_2O)_3$-N | 444 | 2.8 | $d_{15}$=0.72; $d_{24}$=1.27; $d_{33}$=-3.68 | 9.4 | 0.048 |
| $SiGeN_2O$ | *Cm* | cal. [a)] | $(Si/GeN_3)_2$-O | 345 | 3.6 | $d_{11}$=2.50; $d_{24}$=2.16; $d_{33}$=-4.77 | 12.2 | 0.103 |

[a)] Calculated results are from this work. [b)] Related data can also see Ref. [51,52]. [c)] $d_{36}$(KDP)=0.39 pm/V.

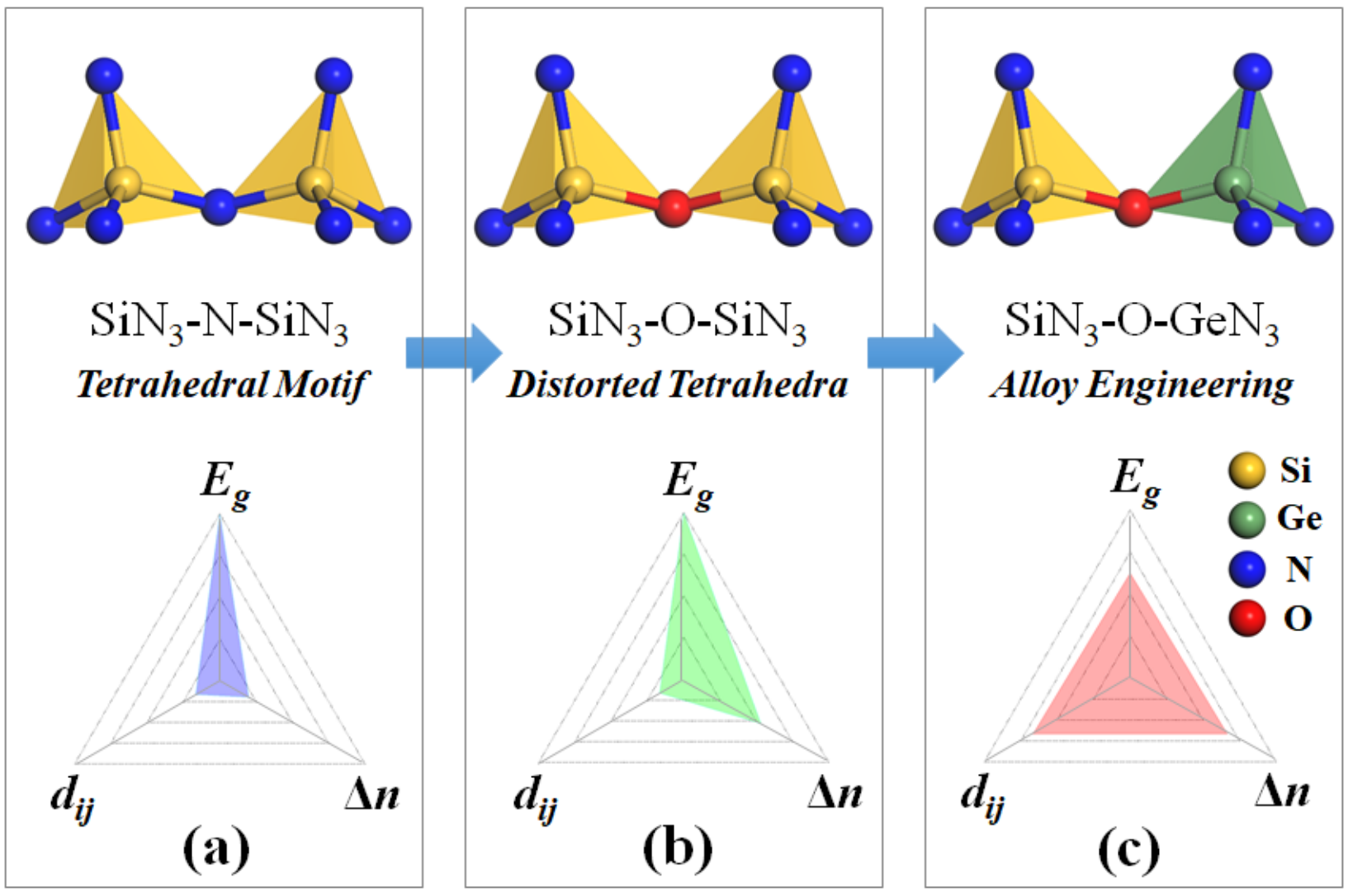

FIG. 1. Design principle for a good balance between the three key NLO parameters. Structural evolution from $SiN_3$-N-$SiN_3$ tetrahedra (a) to $SiN_3$-O-$SiN_3$ tetrahedra (b) and to $SiN_3$-O-$GeN_3$ tetrahedra (c). Their corresponding radar charts of bandgap $E_g$, SHG effect $d_{ij}$ and birefringence $\Delta n$ are shown in the bottom panels.

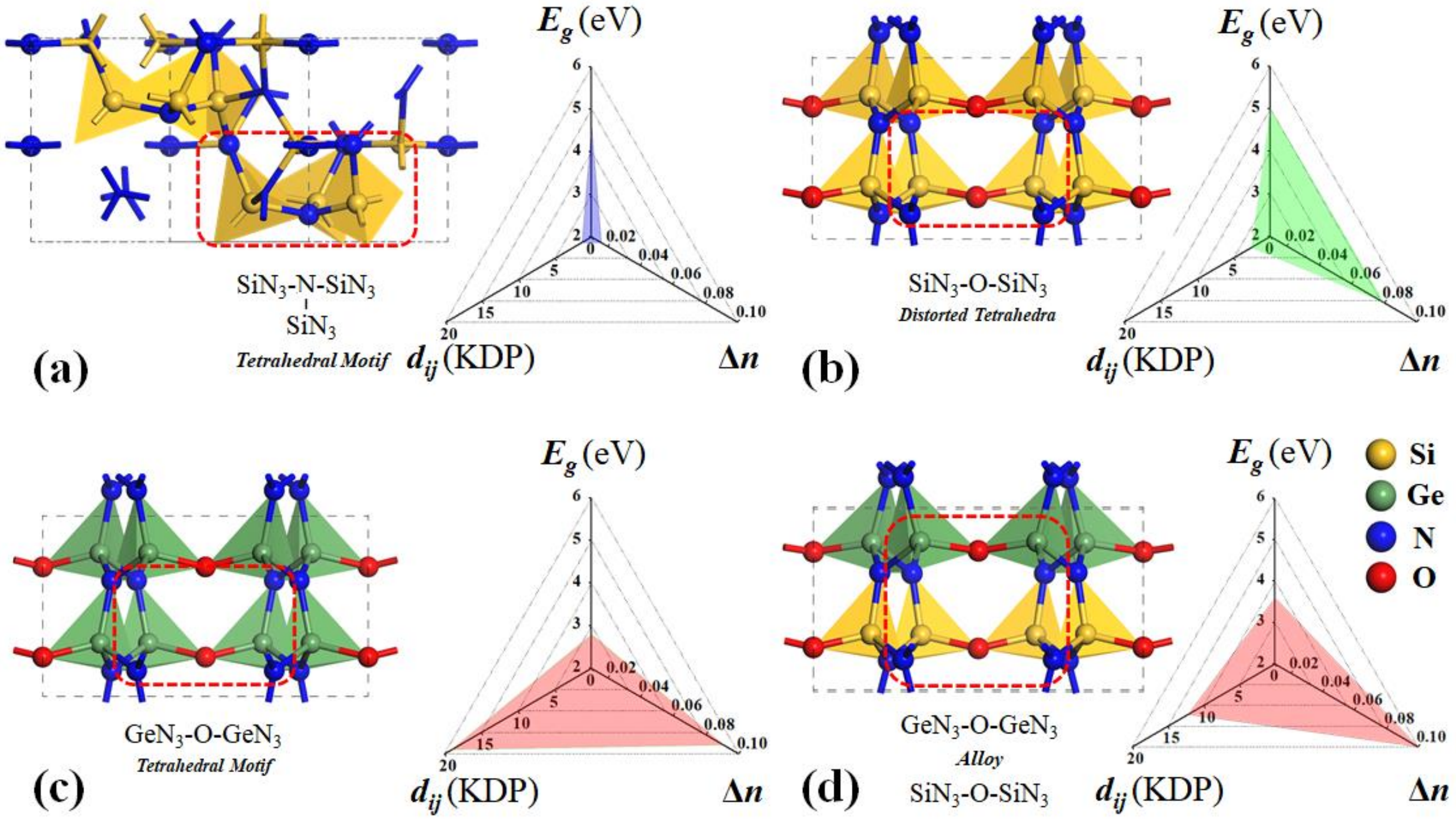

FIG. 2. Crystal structures (left panels) and corresponding radar charts (right panels) of calculated bandgaps $E_g$, SHG effects $d_{ij}$ and birefringence $\Delta n$ in typical silicon nitride α-$Si_3N_4$ (a), silicon oxynitride $Si_2N_2O$ (b), germanium oxynitride $Ge_2N_2O$ (c), and silicon-germanium oxynitride $SiGeN_2O$ (d)

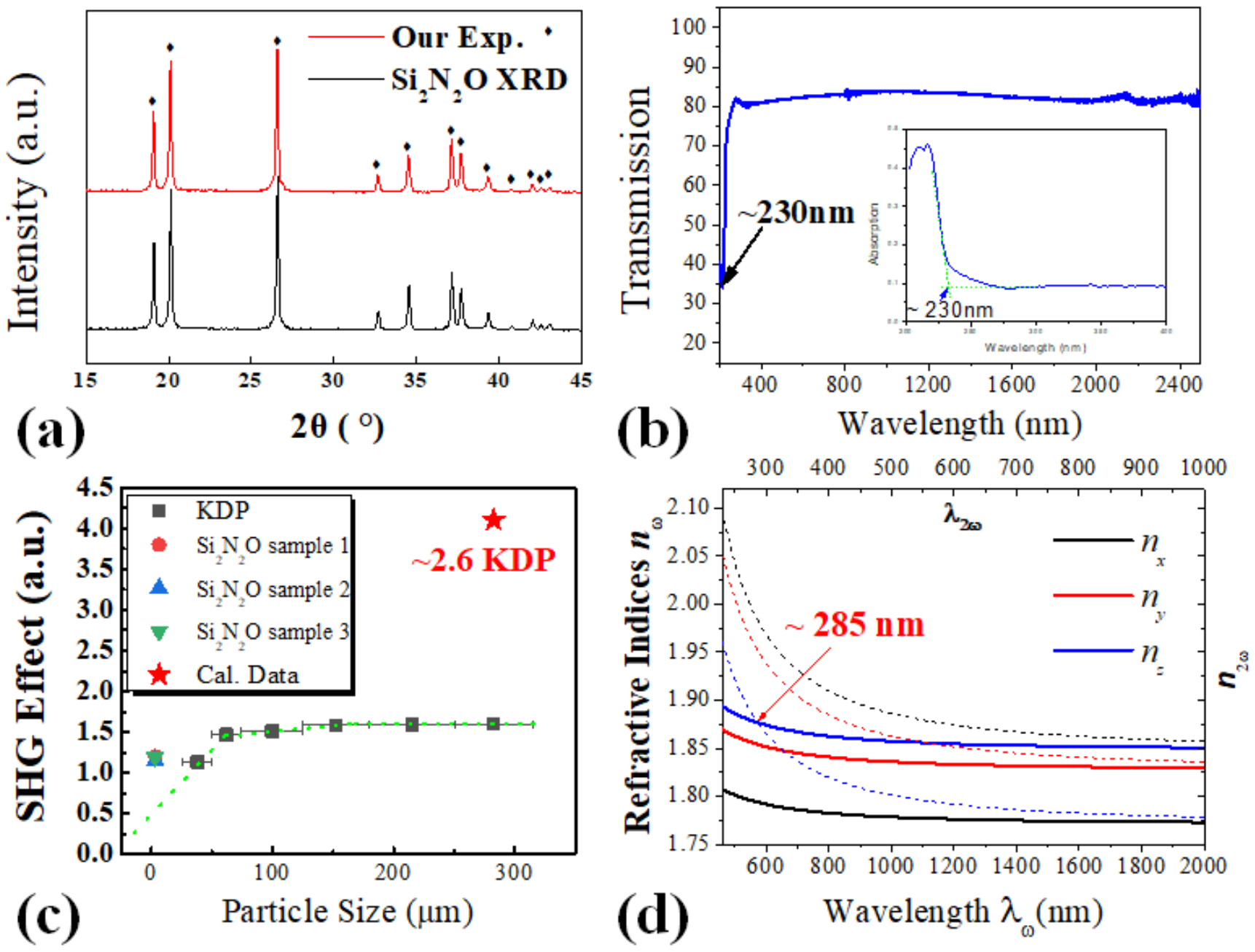

FIG. 3. Experimental measurements of powder XRD (a), UV-visible-IR spectra (b), powder SHG effect (in comparison with KDP) (c), and calculated refractive indices of $Si_2N_2O$ as a function of wavelength (d)

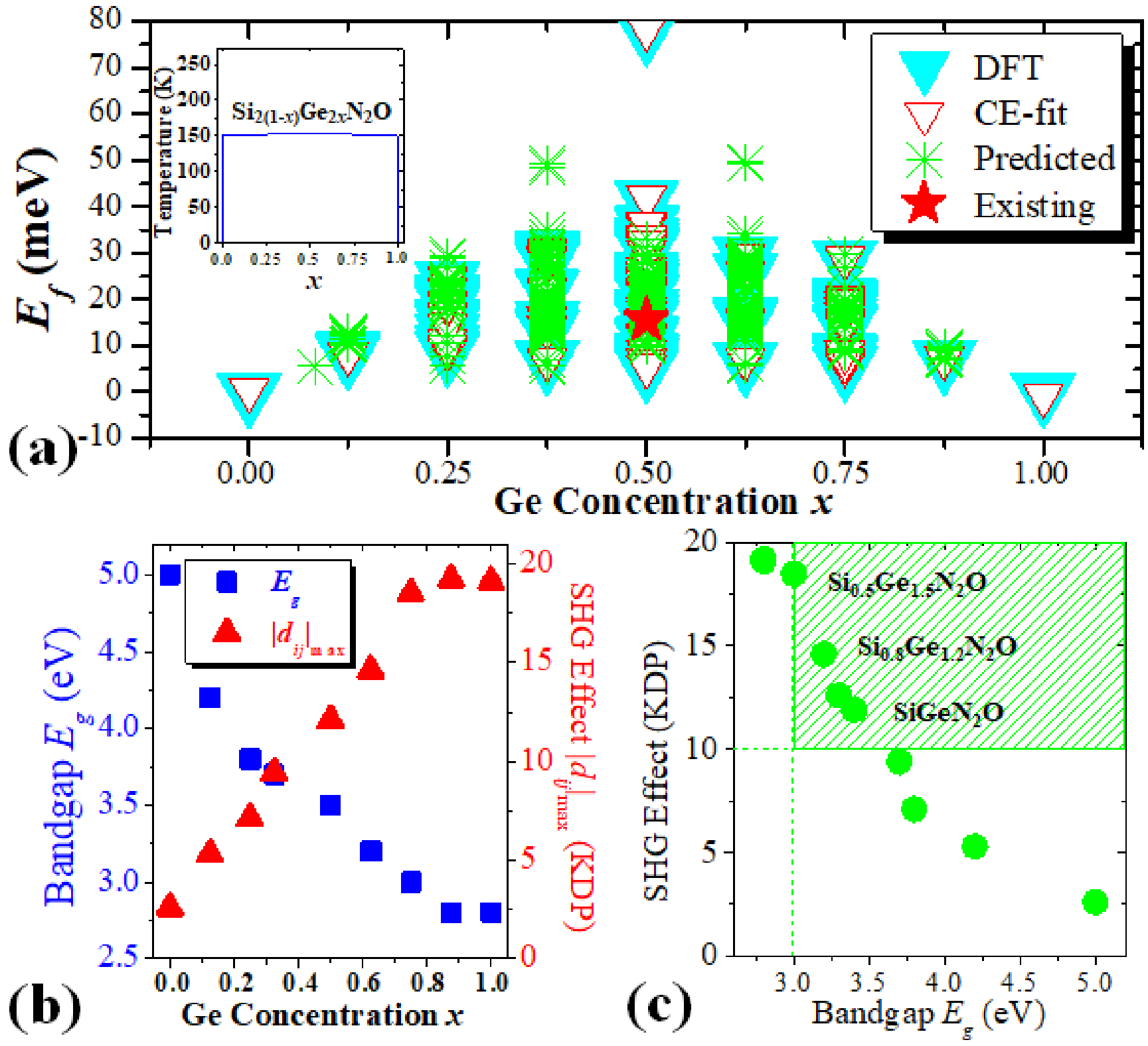

FIG. 4. Calculated formation energies $E_f$ (a), $E_g$ and $d_{ij}$ of $Si_{2(1-x)}Ge_{2x}N_2O$ with respect to the Ge concentration $x$ from 0 to 1 (b), and IR NLO balanced area (c). Red star marked in (a) represent the calculated $E_f$ of experimentally synthesized $SiGeN_2O$. Inset in (a): Monte-Carlo simulated $T$-$x$ phase diagram of $Si_{2(1-x)}Ge_{2x}N_2O$

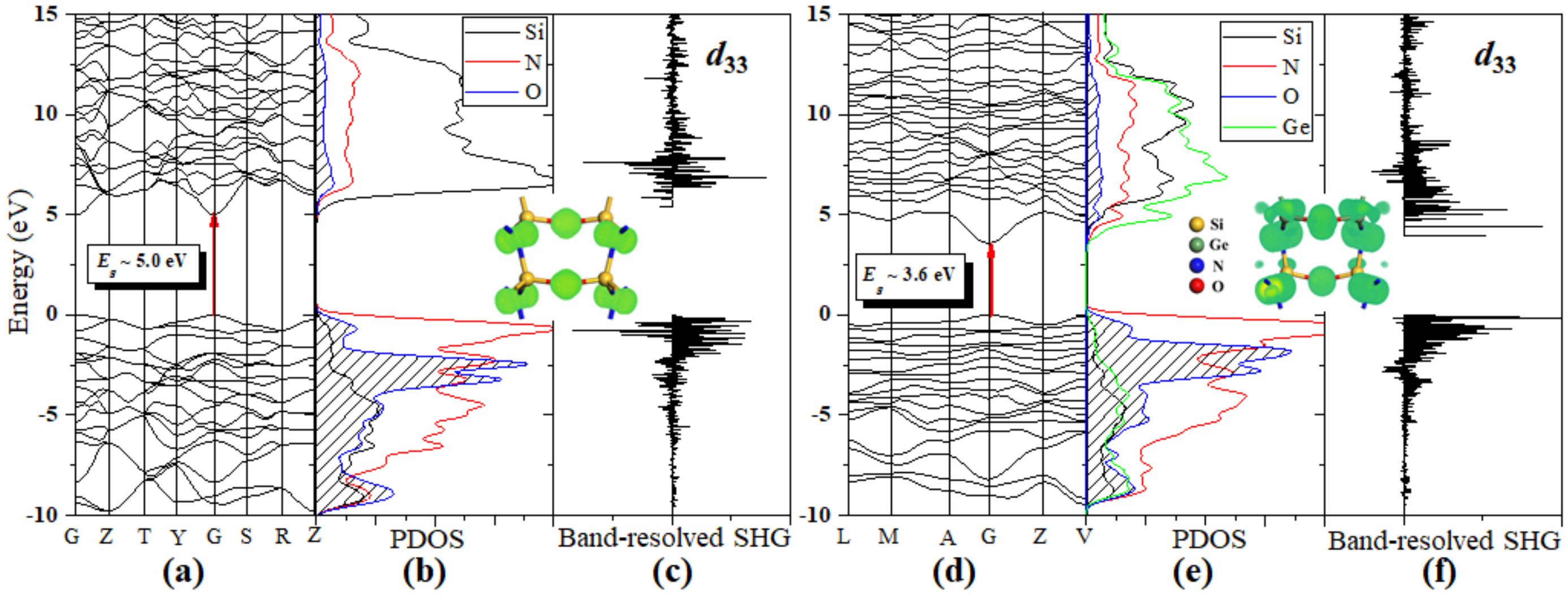

FIG. 5. Calculated band structure (a), PDOS (b), and band-resolved SHG ($d_{33}$) analysis (c) for $Si_2N_2O$. (d-f) are same as (a-c) but for $SiGeN_2O$. Insets are the corresponding SHG density projection for $Si_2N_2O$ and $SiGeN_2O$